\documentclass[runningheads]{llncs}

\AtBeginDocument{%
  \providecommand\BibTeX{{%
    \normalfont B\kern-0.5em{\scshape i\kern-0.25em b}\kern-0.8em\TeX}}}

\usepackage{comment}
\usepackage[utf8]{inputenc}
\usepackage{xcolor}
\usepackage{xspace}
\usepackage{color}
\usepackage{listings}
\usepackage{acronym}
\usepackage[T1]{fontenc}
\usepackage{multirow}
\usepackage{colortbl}
\usepackage{url}
\usepackage{comment}
\usepackage{colortbl}
\usepackage{lipsum}
\usepackage{multicol}
\usepackage{amsmath}

\usepackage{float}
\usepackage{graphicx}
\usepackage{algorithmic}
\usepackage{listings}
\usepackage{amsfonts}
\usepackage{tcolorbox}
\usepackage{wrapfig, blindtext}
\usepackage[linesnumbered, ruled, vlined]{algorithm2e}
\usepackage{longtable}
\usepackage{cancel}
\usepackage{todo}
\usepackage{graphicx}
\usepackage{subcaption} 

\definecolor{lightgray}{rgb}{.9,.9,.9}
\definecolor{darkgray}{rgb}{.4,.4,.4}
\definecolor{darkgreen}{rgb}{0, 0.39, 0.00}
\definecolor{Gray}{gray}{0.7}
\definecolor{codegreen}{rgb}{0,0.6,0}
\definecolor{codegray}{rgb}{0.5,0.5,0.5}
\definecolor{codepurple}{rgb}{0.58,0,0.82}
\definecolor{backcolour}{rgb}{0.95,0.95,0.92}

\lstdefinestyle{mystyle}{
    backgroundcolor=\color{backcolour},   
    commentstyle=\color{codegreen},
    keywordstyle=\color{magenta},
    numberstyle=\tiny\color{codegray},
    stringstyle=\color{codepurple},
    basicstyle=\ttfamily\footnotesize,
    breakatwhitespace=false,         
    breaklines=true,                 
    captionpos=b,                    
    keepspaces=true,                 
    numbers=left,                    
    numbersep=5pt,                  
    showspaces=false,                
    showstringspaces=false,
    showtabs=false,                  
    tabsize=2
}

\acrodef{SAST}{Static Application Security Testing}
\acrodef{AST}{Abstract Syntax Tree}
\acrodef{SFO}{Splunk Forwarder Operator}
\acrodef{HDBSCAN}{Hybrid Density-Based Spatial Clustering of Applications with Noise}
\acrodef{DBSCAN}{Density-Based Spatial Clustering of Applications with Noise}
\acrodef{SME}{Subject Matter Expert}
\acrodef{BERT}{Bidirectional encoder representations from transformers}
\acrodef{APT}{Advanced Persistent Threat}
\acrodef{MST}{Minimum Spanning Tree}
\acrodef{CWE}{Common Weakness Enumeration}
\acrodef{SAST}{Static Application Security Testing}
\acrodef{DAST}{Dynamic Application Security Testing}

\begin{document}

\title{Toward Automated Security Risk Detection in Large Software Using Call Graph Analysis}


\author{Nicholas Pecka$^{1,2}$, Lotfi Ben Othmane$^{1}$, and Renee Bryce$^{1}$}

\institute{\normalsize $^1$University of North Texas, Denton, TX, USA\\
           \normalsize $^2$Red Hat, Kansas City, MI, USA\\
}
\maketitle

\begin{abstract}

Threat modeling plays a critical role in the identification and mitigation of security risks; however, manual approaches are often labor-intensive and prone to error. This paper investigates the automation of software threat modeling through the clustering of call graphs using density-based and community detection algorithms, followed by an analysis of the threats associated with the identified clusters. The proposed method was evaluated through a case study of the Splunk Forwarder Operator (SFO), wherein selected clustering metrics were applied to the software’s call graph to assess pertinent code-density security weaknesses. The results demonstrate the viability of the approach and underscore its potential to facilitate systematic threat assessment. This work contributes to the advancement of scalable, semi-automated threat modeling frameworks tailored for modern cloud-native environments.

\end{abstract}


\section{Introduction}\label{sec:intro}
As the cyber landscape evolves, so does the attack surface of modern applications. A variety of tools exist to uncover vulnerabilities through static and dynamic analysis, but they rarely look beyond the application itself. These tools often ignore how an application interacts with its broader environment, namely deployed infrastructure, integrated components, and system-level dependencies, leading to overlooked threats or irrelevant findings.

Threat modeling employs a system-wide perspective, mapping data flows and interactions from end users to deployed components in order to assess threats applicable to the system. Although typically conducted during the design phase, threat modeling is often neglected after deployment. This presents a significant challenge, as software systems evolve continuously and an initial threat model can quickly become outdated. While vulnerability scanners may identify component-level issues, they rarely capture risks introduced at higher levels of interaction. Moreover, even a single component update (particularly within complex systems) can invalidate an existing model.

Another challenge is that threat modeling remains largely manual and requires input from multiple stakeholders and domain experts. In large systems, no single person (or small group) can realistically maintain a complete, up-to-date model. This complexity often pushes threat modeling down the priority list once the software is in production.

This paper explores automating aspects of the threat modeling process through call graph analysis. Call graphs, generated from existing code, capture both current implementations and future changes. By applying clustering algorithms, we aim to detect structural patterns that signal potential risks. This approach can support threat modeling at scale, ease manual burdens, and enable models to remain relevant throughout the software lifecycle.

Our contributions are as follows:
\begin{enumerate}
\item Evaluation of clustering capabilities of \ac{DBSCAN}~\cite{dbscanOriginal}, \ac{HDBSCAN}~\cite{Malzer_2020}, Louvain~\cite{louvain2025}, and Leiden~\cite{fromLouvainToLeiden}.
\item Proposing an automated approach that applies clustering-based algorithms to call graphs, with heuristics for threat identification.
\item Evaluating the approach on the Splunk Forwarder Operator, a widely used logging agent in Red Hat OpenShift environments.
\end{enumerate}

The remainder of this paper is organized as follows: Section~\ref{sec:relworks} reviews related work; Section~\ref{sec:methods} describes the method; Section~\ref{sec:evaluation} presents results on the Splunk Forwarder Operator; Section~\ref{sec:TtV} outlines threats to validity; and Section~\ref{sec:conclusion} concludes.

\section{Related Work}\label{sec:relworks}
Graph-based techniques and unsupervised learning have been widely applied in cybersecurity, particularly in malware analysis, network defense, and threat intelligence. Although these studies share methods with our work (such as clustering and structural graph analysis) their goals and domains differ. Here, we outline key contributions and how they relate to our focus on proactive, system-wide threat modeling with cluster-based detection.

Herranz-Oliveros et al. apply DBSCAN and \ac{HDBSCAN} to study lateral movement in networks~\cite{adAttackGraph}. Their approach identifies pivot nodes attackers might exploit, offering relevant insights for threat modeling but not aimed at generating a full system model.

Gulbay and Demirici use the Leiden algorithm to analyze \ac{APT} reports~\cite{aptAnalysis}, extracting actionable threat intelligence. Although this aligns with our interest in community detection, their work is reactive (responding to observed attacks), whereas threat modeling seeks to anticipate risks in advance.

Finally, traditional \ac{SAST} and \ac{DAST} tools differ from threat modeling in scope. \ac{SAST} scans code for rule-based vulnerabilities, while \ac{DAST} tests running applications. In contrast, threat modeling evaluates the entire system, components, integrations, and environment, revealing risks beyond code-level flaws.

\section{Call Graph Analysis with Clustering Methods}\label{sec:methods}

To advance automated threat modeling, we turn to call graph analysis. Call graphs capture entry points, data flows, and sensitive interactions that are critical elements for identifying attack surfaces. They also reveal unused or legacy functions that could present hidden risks if left unmonitored. Visualizing these structures allows developers to verify code necessity, trace privileged actions, and better map security-critical pathways.  

\subsection{Clustering Approaches}

We evaluated both density-based and graph-based clustering methods. \ac{DBSCAN}~\cite{dbscanOriginal} and Louvain~\cite{louvain2025} were briefly explored, but left to promote more robust alternatives. Consequently, our primary focus is on \ac{HDBSCAN}~\cite{Malzer_2020} for density-based analysis and Leiden~\cite{fromLouvainToLeiden} for graph-based clustering, both of which demonstrated stronger performance on call graph data.  

\subsection{Density-Based Algorithms}
Density-based clustering identifies regions of high connectivity while marking sparse nodes as noise, making it effective for anomaly detection. A common evaluation metric is the silhouette score~\cite{silhouettesClusterAnalysis}, defined as:  

\[
silhouette =  \frac{b-a}{\max(a,b)}
\]  


where \(a\) is the average distance from a point to others in the same cluster (intra-cluster distance) and \(b\) is the average distance from a point to those in the nearest neighboring cluster (where the nearest neighboring cluster is found by the smallest average distance of a cluster to the selected point). Scores near 1 indicate well-defined clusters, while scores near 0 or negative values suggest weak assignments or misclassifications. Averaging across all points yields a global measure of clustering quality, especially valuable when ground truth labels are unavailable.  

While DBSCAN~\cite{dbscanOriginal} provided a baseline, its reliance on a fixed $\varepsilon$ value limited its utility in non-uniform call graphs.  

\ac{HDBSCAN}~\cite{Malzer_2020} removes this limitation by considering variable densities and forming a hierarchy of clusters. It introduces the concept of mutual reachability distance:  

\[
\text{mutual\_reachability}(a, b) = \max\left\{\text{core}_k(a),\ \text{core}_k(b),\ d(a, b)\right\}
\]  

where \(d(a, b)\) is the distance between points \(a\) and \(b\), \(\text{core}_k(a)\) is the distance from \(a\) to its $k^{\text{th}}$ nearest neighbor, and \(k\) is the minimum number of neighbors defining a dense region. This transformation enables construction of a minimum spanning tree, from which stable clusters are extracted across density levels. For call graph analysis, this flexibility highlights dense, interconnected components while isolating sparse or anomalous regions (potential signals of risk).

\subsection{Graph-Based Algorithms}
Graph-based clustering identifies communities in which nodes are more densely connected internally than externally. The modularity metric~\cite{modularityCommunityStructure} evaluates the quality of such partitions:  

\[
Q = \frac{1}{2m} \sum_{i,j} \left( A_{ij} - \frac{k_i k_j}{2m} \right) \delta(c_i, c_j)
\]  

where \(A_{ij}\) is the edge weight between nodes \(i\) and \(j\), \(k_i = \sum_j A_{ij}\) is the degree of node \(i\), \(m = \frac{1}{2} \sum_{i,j} A_{ij}\) is the total edge weight, \(c_i\) is the community of node \(i\), and \(\delta(c_i, c_j)\) is 1 if \(c_i = c_j\) and 0 otherwise.  

Louvain~\cite{improvingLouvainCommunityDetection} was initially tested but produced inconsistent partitions for our dataset.  

The Leiden method~\cite{fromLouvainToLeiden} improves upon Louvain by introducing a refinement phase that ensures each community is internally connected:  

\[
\forall u, v \in C_k,\ \exists \text{ a path } u \leftrightarrow v \text{ such that } (u,v) \in E(C_k)
\]  

This guarantees meaningful communities before aggregation, avoiding unstable or fragmented clusters. For call graphs, Leiden provides more reliable, convergent results, capturing trust boundaries and structural patterns critical for system-wide threat modeling.

\section{Evaluation}\label{sec:evaluation}

\subsection{Case Study: Splunk Forwarder Operator}

The \ac{SFO}~\cite{sfo2025} is a Red Hat~\cite{redhat2022} operator deployed by default on OpenShift~\cite{openshift2023} to collect logs from the cluster and forward them to an external Splunk instance~\cite{splunk2025}. Its widespread use across real-world customer clusters makes it an ideal case for evaluating automated threat modeling, as it encounters diverse runtime environments and integration patterns.

For analysis, we used \texttt{go-callvis}~\cite{go-callvis} to perform static analysis and generate a complete call graph from the operator’s \texttt{main.go} file. This graph captures internal function calls and control flow, providing the structural representation necessary for clustering-based threat detection.

The call graph serves as input to the clustering algorithms (\ac{DBSCAN}, \ac{HDBSCAN}, Louvain, and Leiden), allowing us to assess how well each method isolates outliers, identifies modular structures, and highlights potential threats across the operator’s architecture.

\subsection{Algorithm Comparison utilizing Call Graphs}

To evaluate clustering algorithms on large-scale call graphs, we generated a graph from the \texttt{main.go} file of \ac{SFO} using Go tooling, i.e., \texttt{go-callvis}  The resulting graph comprised over 350,000 lines of call data, representing 39,024 nodes and 286,302 edges.

We applied \ac{DBSCAN}, \ac{HDBSCAN}, Louvain, and Leiden to assess their performance. Density-based methods were evaluated using the silhouette score, while graph-based methods used modularity. This comparison highlights the relative strengths of each approach and guides selection for deeper analysis.

\textbf{DBSCAN} struggled at this scale. The best configuration (\texttt{eps}=0.09, minimum sample size=5) produced 32 clusters in 2.64 seconds with a silhouette score of only 0.0227, indicating near-random clustering.

\begin{figure}[bth]
    \centering
    \begin{subfigure}[b]{0.48\textwidth}
        \centering
        \includegraphics[width=\textwidth]{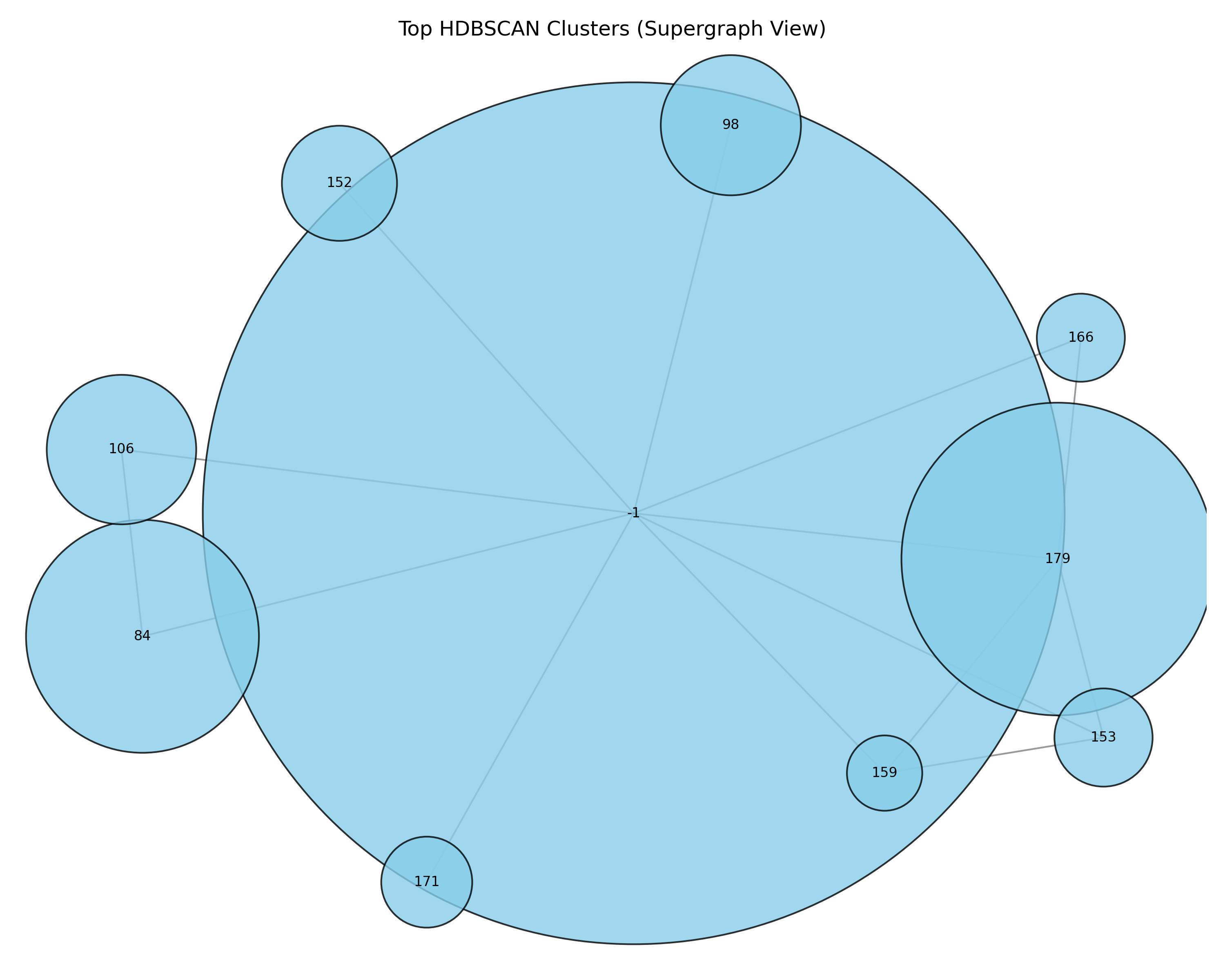}
        \caption{Top 10 cluster illustration of \ac{HDBSCAN} algorithm for \ac{SFO}.}
        \label{fig:hdbscan10}
    \end{subfigure}
    \hfill
    \begin{subfigure}[b]{0.48\textwidth}
        \centering
        \includegraphics[width=\textwidth]{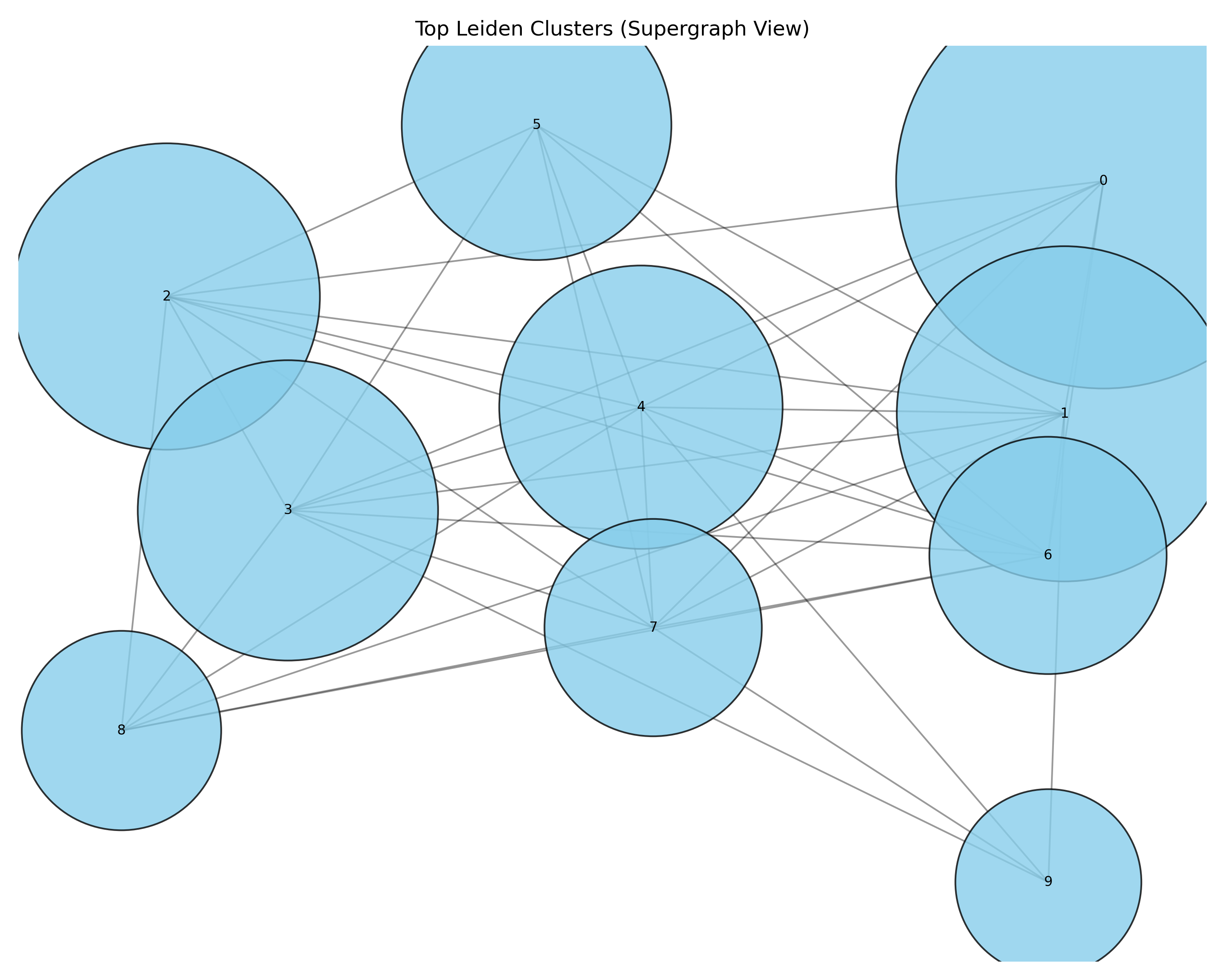}
        \caption{Top 10 cluster illustration of Leiden algorithm for \ac{SFO}.}
        \label{fig:leiden10}
    \end{subfigure}
    \caption{Comparison of top 10 clusters produced by \ac{HDBSCAN} and Leiden algorithms for \ac{SFO}.}
    \label{fig:clusters10}
\end{figure}

\textbf{HDBSCAN} addressed DBSCAN’s limitations, handling noise and varying cluster density effectively. With a minimum cluster size of 8, it produced 180 clusters in 18.17 seconds with a silhouette score of 0.5087, demonstrating meaningful structure well above random behavior (see Figure~\ref{fig:hdbscan10}).

\textbf{Louvain} generated 367 clusters in 6.45 seconds with a modularity score of 0.8090, successfully identifying modular structures in the call graph.

\textbf{Leiden} further improved on Louvain, producing 366 clusters in just 0.26 seconds with a slightly higher modularity of 0.8202. Its refinement phase produced higher quality clusters and drastically improved runtime, making it especially suitable for iterative or large-scale threat modeling (see Figure~\ref{fig:leiden10}).

These results provide a clear basis for selecting \ac{HDBSCAN} and Leiden for more detailed threat analysis in the following sections.

\subsection{Heuristics for Threat Identification}

With the initial evaluation of clustering algorithms complete, we shift focus towards analyzing the clustering output for potential threat vectors. While the clustering results provide structure to an otherwise massive and complex graph, identifying security concerns within those structures requires additional analysis. To move from structural insights to actionable threat intelligence, we apply a set of heuristics designed to flag patterns commonly associated with insecure or suspicious behavior in software systems based on the MITRE corporations \ac{CWE} category system~\cite{cwe2025} relevant to software complexity. Table~\ref{tab:complexitythreats} provides the threats with a citation that inspired the heuristic, their detailed descriptions, and an explanation of how they are associated with \acp{CWE}.

\begin{table}[h]
        \centering
          \caption{Threats related to software complexity and associated weaknesses.}
\label{tab:complexitythreats}  
\scalebox{.8}{\begin{tabular}{|p{1.3in}|p{2.6in}|p{2.0in}|}

\hline 
\rowcolor{gray!18}
 \begin{center}
\textbf{{\small Threat}}
\end{center}
& \begin{center}
\textbf{{\small Description}}
\end{center}
& \begin{center}
\textbf{{\small Associated Potential Weakness}}
\end{center}
\\ \hline

 Small clusters with many external connections (bridging)~\cite{smallClusterWithManyOthers} & These clusters act as bridges between multiple larger clusters and may serve as entry points or intermediaries that cross trust boundaries. & This behavior aligns with \ac{CWE} 668~\cite{cwe668}: Exposure of Resource to Wrong Sphere. \\ \hline
    
 Large clusters with high incoming call volume (hotspots)~\cite{cunha2015moduleAttack},~\cite{stergiopoulos2015centrality} & Clusters that receive a large number of incoming calls may act as critical processing units or bottlenecks. These can represent central services or APIs and could be likely targets for access control violations. & We associate this heuristic with \ac{CWE} 284~\cite{cwe284}: Improper Access Control.
    \\ \hline
    
  Dangling nodes~\cite{danglingNodes} & Nodes that only connect to a single node in a different cluster and have no internal cluster connectivity may represent loosely controlled logic or misplaced code. The additional outlying logic could indicate a potential injection or code that wasn't properly cleaned up during the software development life cycle.& We associate this heuristic with \ac{CWE} 94~\cite{cwe94}: Improper Control of Generation of Code ('Code Injection') or \ac{CWE} - 1164~\cite{cwe1164}: Irrelevant Code. \\ \hline
    
Hub nodes~\cite{hubNodes} & Individual nodes with an unusually high number of connections, either incoming or outgoing, could indicate points of aggregation, routing, or parsing. & These hubs often correlate with \ac{CWE} 20~\cite{cwe20}: Improper Input Validation, especially if they process untrusted data without sufficient sanitization.\\ \hline
    
Weak clusters with low internal edge ratios~\cite{weakClustersLowInternalEdgeRatio} & Clusters that have more connections leaving than staying within may indicate poor encapsulation or broken modularity, often exposing internal data or logic externally. & Due to these we associate the heuristic with \ac{CWE} 200~\cite{cwe200}: Exposure of Sensitive Information to an Unauthorized Actor.\\ \hline
\end{tabular} }%
\end{table}

We use this initial result to assess potential threats of the evaluated software. We plan to conduct a comprehensive literature review to identify the association between cluster types and potential threats, which will allow for exhaustive assessments of potential threats for large cloud-based software. 

\begin{table}[h]
        \centering
          \caption{Heuristic results of associating the clusters generated using HDBSCAN and Leiden methods from SFO to software density-based \ac{CWE}.}
\label{tab:heuristicResults}  
\scalebox{.8}{\begin{tabular}{|p{1.5in}|p{1.0in}|p{1.0in}|p{1.0in}|}

\hline 
\rowcolor{gray!18}
 \begin{center}
\textbf{{\small Heuristic}}
\end{center}
& \begin{center}
\textbf{{\small HDBSCAN}}
\end{center}
& \begin{center}
\textbf{{\small Leiden}}
\end{center}
& \begin{center}
\textbf{{\small Weakness}}
\end{center}
 \\
\hline
Bridging Clusters& 0& 0& CWE - 668\\
\hline
Hotspot Clusters& 162& 24& CWE - 284\\
\hline
Dangling Nodes& 57& 485& CWE - 123\\
\hline
Hub Nodes& 80& 217& CWE - 20\\
\hline
Weak Clusters& 27& 0& CWE -200\\
\hline
\rowcolor{lightgray!18}
Total number of clusters &180 &366 & \\
\hline

\end{tabular} }%
\end{table}

\subsection{Findings and Observations}

For the heuristic-driven threat detection analysis, we focused specifically on the output of the \ac{HDBSCAN} and Leiden algorithms, given their strong performance in the previous sections clustering evaluation. For the scope of this paper, we examine the top three results for each heuristic applied to both algorithms. It is important to preface that the results of these heuristics are naturally correlated with quality metrics (silhouette score for \ac{HDBSCAN} and modularity for Leiden) as discussed earlier. With additional parameter tuning, more precise or distinct findings may emerge, which we consider for future work.

\noindent\textbf{HDBSCAN} yielded actionable insights for \textit{four out of the five} heuristics, as shown in Table~\ref{tab:heuristicResults}. The heuristics flagged different clusters, each offering a lens into possible security concerns or tuning opportunities. For instance, both the \textit{dangling node} and \textit{hub node} heuristics each returned top results with similar neighbor or connection ratios, suggesting that some of these may be noise rather than clear signals. Expanding the evaluation to more candidates would help determine whether these clusters warrant investigation or are artifacts of imperfect partitioning.

More promising are the results from the \textit{hotspot} heuristic. Cluster 179 was flagged with a significant call volume of 8,338, while the second highest, cluster 98, had 6,806. This level of disparity suggests a disproportionate concentration of traffic in a few clusters, which may warrant prioritization in security assessments.

The most compelling results came from the \textit{weak cluster} heuristic. The top-ranked cluster (Cluster 152) had 458 nodes with an external-to-internal edge ratio of 687, whereas the next result, had only 24 nodes with a ratio of 120. This steep drop-off in both size and edge ratio indicates that Cluster 152 is significantly more exposed or loosely coupled and may represent a potential risk.

\noindent\textbf{Leiden}, on the other hand, reported findings for \textit{three out of the five} heuristics (see Table~\ref{tab:heuristicResults}). A noteworthy observation across multiple heuristics was the consistent appearance of the node \texttt{(reflect.Value).Call}~\cite{goReflect2025}. This function surfaced as the top candidate in both the \textit{dangling} and \textit{hub} heuristics. It exhibited a connection split of 1,968 outgoing calls to different clusters out of a total of 3,447, while the second-ranked node had a mere 64/65 split. Furthermore, \texttt{(reflect.Value).Call} was identified as a hub across 23 clusters, compared to only 9 clusters for the second-ranked hub node.

This suggests poor cohesion and potential misuse or overuse of the \texttt{reflect} package. In Go, the \texttt{reflect} package enables runtime manipulation of types and values, which can bypass compile time checks, introduce runtime instability, and hinder debugging. Its ability to modify addressable values adds another layer of risk, particularly in a high-usage context like the one observed in this study. These characteristics may offer an attacker more flexibility and stealth, making this node a prime candidate for further investigation.

These heuristic findings represent the current endpoint of our research. They provide a filtered and structured view of potentially vulnerable areas in the application, setting the stage for deeper manual analysis in future work. Our next steps include evaluating the flagged clusters for specific vulnerabilities and validating these patterns against known security flaws.

\section{Threats to Validity}\label{sec:TtV}

This research is an early-stage exploration. Its limited scope may affect the generalization of our results.

First, the study relies on a single case—the \ac{SFO} operator. While widely deployed, tuning clustering algorithms on this one application may introduce bias. Parameters and heuristics may not be generalized to other software systems with different architectures, sizes, or domain-specific behaviors.

Second, although \ac{HDBSCAN} and Leiden showed promising results, we have not yet tested them on significantly larger or more fragmented call graphs. Future work should apply these methods across a broader set of applications and stress-test them under diverse structural conditions.

Finally, the heuristics for identifying potential threats are based on common vulnerability patterns (e.g., bridging, hotspots, weak cohesion) and \ac{CWE} mappings. They have not been empirically validated or reviewed by experts to confirm their predictive value for real-world vulnerabilities. We plan to address this in future work.

Despite these limitations, the findings highlight a promising path forward. Future iterations will focus on refining algorithms, expanding validation, and improving practical applicability.

\section{Conclusion}\label{sec:conclusion}

This paper explores automating threat modeling, with a long-term vision of building a fully automated pipeline. We evaluated four clustering algorithms, \ac{DBSCAN}, \ac{HDBSCAN}, Louvain, and Leiden, on a large-scale call graph generated from the \ac{SFO}, a production-grade operator widely used in OpenShift environments.

Comparing algorithms by their natural strengths (silhouette score for density-based methods, modularity for graph-based methods), we found \ac{HDBSCAN} and Leiden to be the most effective. \ac{HDBSCAN} produced meaningful clustering with a silhouette score of 0.5087 and 180 clusters, while Leiden achieved a modularity score of 0.8202 and outperformed Louvain in runtime.

Applying our heuristics revealed weaknesses such as poor cohesion, excessive external communication, and hub-like behavior. This helped surface critical nodes like \texttt{(reflect.Value).Call}, which might otherwise remain hidden in the complexity of the graph, demonstrating the value of combining clustering with targeted analysis.

Ultimately, this work bridges static program analysis with intelligent graph clustering, laying a foundation for scalable, automated threat modeling in real-world systems.

\section*{Acknowledgment}\label{sec:acknowledgement}
ChatGPT was used to check and improve the spelling and grammar of the paper and improve the flow of paragraphs of some sections.

\RestyleAlgo{ruled}
\SetKwComment{Comment}{/* }{ */}

\bibliography{sample}

\bibliographystyle{unsrt}


\end{document}